\documentclass[aps,amsmath,twocolumn,amssymb,floatfix,showpacs,superscriptaddress,nofootinbib,longbibliography]{revtex4-1}
\usepackage{MnSymbol}
\usepackage{braket}
\usepackage[dvipsnames]{xcolor}
\usepackage{float}
\usepackage{subfigure}
\usepackage{tikz}
\usepackage[colorlinks=true,linktoc=page,linkcolor=red,citecolor=blue,urlcolor=purple]{hyperref}

\newcommand*\circled[1]{\tikz[baseline=(char.base)]{\node[shape=circle,draw,inner sep=1.0pt] (char) {#1};}} 
\newcommand{\vect}[1]{\boldsymbol{\mathrm{#1}}}
\newcommand{\Eq}[1]{Eq.~(\ref{#1})}
\mathchardef\mhyphen="2D 

\newcommand{\ie}{{ i.e.,\,\,}}

\newcommand{\ua}{{\uparrow }}
\newcommand{\da}{{\downarrow }}

\newcommand\bea{\begin{eqnarray}}
\newcommand\eea{\end{eqnarray}}
\newcommand\beq{\begin{equation}}  
\newcommand\eeq{\end{equation}}

\newcommand{\non}{\nonumber}  
\usepackage[normalem]{ulem}
\definecolor{lime}{HTML}{A6CE39}
\usepackage{sidecap,tikz}
\DeclareRobustCommand{\orcidicon}{\hspace{-1.0mm}
	\begin{tikzpicture}
		\draw[lime, fill=lime] (0.0,0.0) 
		circle [radius=0.15] 
		node[white] {{\fontfamily{qag}\selectfont \tiny \,ID}};
		\draw[white, fill=white] (-0.0525,0.095) 
		circle [radius=0.007];
	\end{tikzpicture}
	\hspace{-3.0mm}
}
\foreach \x in {A, ..., Z}{\expandafter\xdef\csname orcid\x\endcsname{\noexpand\href{https://orcid.org/\csname orcidauthor\x\endcsname}{\noexpand\orcidicon}}
}

\AtBeginDocument{%
	\newwrite\bibnotes
	\def\bibnotesext{Notes.bib}
	\immediate\openout\bibnotes=\jobname\bibnotesext
	\immediate\write\bibnotes{@CONTROL{REVTEX41Control}}
	\immediate\write\bibnotes{@CONTROL{%
			apsrev41Control,author="08",editor="1",pages="1",title="1",year="1"}}
	\if@filesw
	\immediate\write\@auxout{\string\citation{apsrev41Control}}%
	\fi
}%

\begin{document}

\title{Time evolution of Majorana corner modes in Floquet second-order topological superconductor}  

\author{Arnob Kumar Ghosh\orcidA{}}
\email{arnob@iopb.res.in}
\affiliation{Institute of Physics, Sachivalaya Marg, Bhubaneswar-751005, India}
\affiliation{Homi Bhabha National Institute, Training School Complex, Anushakti Nagar, Mumbai 400094, India}

\author{Tanay Nag\orcidB{}}
\email{tanay.nag@physics.uu.se}
\affiliation{Department of Physics and Astronomy, Uppsala University, Box 516, 75120 Uppsala, Sweden}

\author{Arijit Saha\orcidC{}}
\email{arijit@iopb.res.in}
\affiliation{Institute of Physics, Sachivalaya Marg, Bhubaneswar-751005, India}
\affiliation{Homi Bhabha National Institute, Training School Complex, Anushakti Nagar, Mumbai 400094, India}

\begin{abstract}
We propose a practically feasible time-periodic sinusoidal drive protocol in onsite mass term to generate the two-dimensional~(2D) Floquet second-order topological superconductor, hosting both the 
regular $0$- and anomalous $\pi$-Majorana corner modes~(MCMs) while starting from a static 2D topological insulator/$d$-wave superconductor heterostructure setup. We theoretically study the local density spectra and the time dynamics of MCMs in the presence of such drive. The dynamical MCMs are topologically characterized by employing the average quadrupolar motion. Furthermore, we employ the Floquet perturbation theory~(FPT) in the strong driving amplitude limit to provide analytical insight into the problem. We compare our exact (numerical), and the FPT results in terms of the eigenvalue spectra and the time dynamics of the MCMs. We emphasize that the agreement between the exact numerical and the FPT results are more prominent in the higher frequency regime for close to the $0$-quasi-energy mode. 
\end{abstract}

\maketitle

\section{Introduction}
The quest for first-order topological superconductors~(TSCs) has been engaging substantial attention among the quantum condensed matter physics community for the past two decades. These TSCs have become the breeding ground for realizing the Majorana zero-modes~(MZMs)~\cite{Kitaev_2001,qi2011topological,Alicea_2012}. These MZMs are predicted to be associated with one-dimensional~(1D) spinless $p$-wave superconductor as proposed by Kitaev~\cite{Kitaev_2001}. From the practical perspective, they can also be perceived in a heterostructure setup consisting of a 1D Rashba nanowire with strong spin-orbit coupling~(SOC) and proximity coupled to an $s$-wave superconductor~\cite{SauPRL2010,LutchynPRL2010,HaimPRL2015}. Interestingly, such MZMs follow non-Abelian statistics, owing to their non-local properties, and can be the building blocks for fault-tolerant quantum computers~\cite{Ivanov2001,freedman2003topological,KITAEV20032,NayakRMP2008}. The hunt for the MZMs is not limited to the theoretical proposals only, and there have been a few experimental advancements~\cite{das2012zero,Deng1557,Mourik2012Science,NichelePRL2017,Zhang2017NatCommun,He294,Grivnin2019,ChenPRL2019} in this direction. However, a distinct signature of the MZMs is yet to be discovered. Moving our attention towards the paradigm of higher-order topological insulators~(HOTIs)~\cite{benalcazar2017,benalcazarprb2017,Song2017,Langbehn2017,schindler2018,Franca2018,wang2018higher,Ezawakagome,Roy2019,Trifunovic2019,Khalaf2018,Szumniak2020,Ni2020,BiyeXie2021,
trifunovic2021higher} and their superconducting counterparts, \ie higher-order topological superconductors~(HOTSCs)~\cite{Geier2018,Zhu2018,Liu2018,Yan2018,WangWeak2018,ZengPRL2019,Zhang2019,ZhangFe2019PRL,Volpez2019,YanPRB2019,Ghorashi2019,GhorashiPRL2020,Wu2020,jelena2020HOTSC,
BitanTSC2020,SongboPRR12020,SongboPRR22020,SongboPRB2020,kheirkhah2020vortex,PlekhanovPRB2020,ApoorvTiwari2020,YanPRL2019,AhnPRL2020,luo2021higherorder2021,QWang2018,
Ghosh2021PRB,RoyPRBL2021,FuBoPRBL2021,Ghosh2022NHPRBL}, 
they are characterized by the presence of $(d-n)$-dimensional electronic~(Majorana) boundary modes; with $d$ being the dimension of the system and $n$~($d \ge n \ge 2$) stands for the order of the topological system. There have been a few concurrent experimental developments to realize the HOTI phase in solid-state systems~\cite{schindler2018higher,Experiment3DHOTI.VanDerWaals}, phononic crystals~\cite{serra2018observation}, acoustic systems~\cite{xue2019acoustic,ni2019observation,Experiment3DHOTI.aSonicCrystals}, electric-circuit setups~\cite{imhof2018topolectrical}, photonic lattice~\cite{PhotonicChen,PhotonicXie} etc.

\begin{figure}[]
	\centering
	\subfigure{\includegraphics[width=0.4\textwidth]{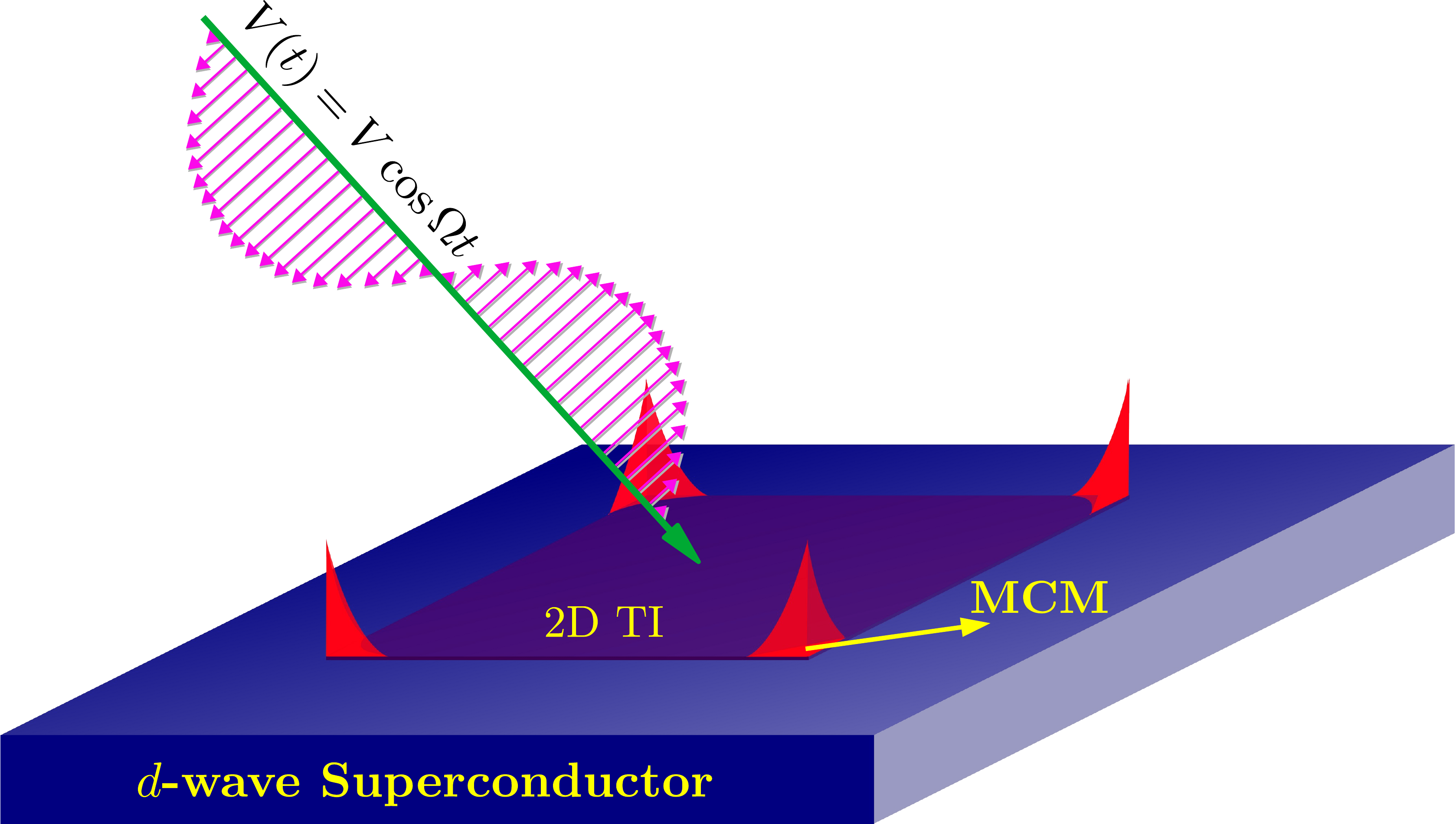}}
	\caption{We depict the schematic representation of our setup, which consists of a two-dimensional~(2D) topological insulator~(TI) in proximity to a $d$-wave superconductor while an external 
	time-dependent 
	periodic potential is harmonically driving the system to realize the $0$ and $\pi$-Floquet Majorana corner modes~(MCMs). 
	}
	\label{static}
\end{figure}

On the other hand, the realm of non-equilibrium systems involving Floquet generation of topological bands provides us with the on-demand control of the topological properties of a system~\cite{kitagawa11transport,lindner11floquet,Rudner2013,Usaj2014,Piskunow2014,Eckardt2015,Eckardt2017,Yan2017,oka2019,NHLindner2020,UmerPRB2020,nag2021anomalous}. Floquet engineering enables one to 
achieve topological phase from a topologically trivial system. 
In a driven system, one can realize the regular $0$- as well as the so-called anomalous $\pi$-modes which do not have any static analog. The latter have been proposed to exist in the first-order Floquet topological systems~\cite{Rudner2013,Usaj2014,Piskunow2014,Eckardt2017,Yan2017}. A few experimental proposals also come up with the brisk theoretical developments~\cite{WangScience2013,McIver2020,Peng2016,fleury2016floquet,RechtsmanExperiment2013,Maczewsky2017}. Very recently, the concept of Floquet engineering has been extended to generate the 
Floquet HOTI~(FHOTI)~\cite{Bomantara2019,Nag19,YangPRL2019,Seshadri2019,Martin2019,Ghosh2020,Huang2020,HuPRL2020,YangPRR2020,Nag2020,ZhangYang2020,bhat2020equilibrium,chaudharyphononinduced2020,
GongPRBL2021,JiabinYu2021,Vu2021,ghosh2021systematic,du2021weyl,WuPRBL2021,ning2022tailoring} and the Floquet HOTSC (FHOTSC)~\cite{PlekhanovPRR2019,BomantaraPRB2020,RWBomantaraPRR2020,ghosh2020floquet,ghosh2020floquet2,VuPRBL2021,GhoshDynamical2022}. The FHOTI has been experimentally contextualized in a meta-material platform, specifically in the acoustic system in Ref.~\cite{experimentFloquetHOTI}. However, a real material-based experiment to realize the FHOTI/FHOTSC phase has not been reported so far, to the best of our knowledge.

In the current literature, there exists a few proposals, relying upon the periodic kick or step-drive protocol to realize the FHOTI and FHOTSC phases hosting only the $0$-modes~\cite{Seshadri2019,bhat2020equilibrium,Nag19,Nag2020,ghosh2020floquet,ghosh2020floquet2} as well as both the regular $0$- and anomalous $\pi$-modes~\cite{Bomantara2019,Martin2019,Huang2020,HuPRL2020,GongPRBL2021,JiabinYu2021,Vu2021,ghosh2021systematic,GhoshDynamical2022}. A handful schemes exist to generate the FHOTI or FHOTSC phase using periodic sinusoidal~(harmonic) driving protocol~\cite{YangPRL2019,chaudharyphononinduced2020,PlekhanovPRR2019,RWBomantaraPRR2020,VuPRBL2021} and periodic laser irradiation~\cite{Ghosh2020,ZhangYang2020,ning2022tailoring}. The previous studies emphasized on the generation of both regular $0$- and anomalous $\pi$-modes in the FHOTI/FHOTSC phase that are primarily based on sinusoidal temporal variation of the hoppings~\cite{YangPRL2019,chaudharyphononinduced2020,RWBomantaraPRR2020,VuPRBL2021}. In Ref.~\cite{PlekhanovPRR2019}, an oscillating magnetic field is considered to capture the FHOTI phase; however, the anomalous $\pi$-mode has not been investigated in this model.
On the other hand, periodic sinusoidal/laser irradiation mediated  FHOTI/FHOTSC phase, hosting both $0$- and $\pi$-modes, is technically not as straightforward as the periodic step/kick drive. Due to the continuous time-sequence nature of the above protocols, one cannot obtain the Floquet operator in a closed analytical form. Finding the parameter space where one can realize the anomalous modes 
is thus cumbersome. In recent times, a few reports have been put forward where a perturbative scheme based on the Floquet perturbation theory~(FPT), has been employed to study a driven system in the strong driving limit~\cite{Sen2021,Mukherjee2020,Sengupta2022Weyl}. Following this background, we pose the following intriguing questions- (1) How to engineer the two-dimensional~(2D) Floquet second-order topological superconductor~(FSOTSC) hosting both $0$- and $\pi$-Majorana corner modes~(MCMs) employing a periodically varying onsite mass term that can be practically more feasible? 
(2) How does these MCMs evolve with time? (3) How to topologically characterize these modes using a proper dynamical invariant? (4)  Can the numerical results, in the strong driving limit, be understood using the analytical FPT approach?  We intend to investigate these interesting questions in this article that are unanswered so far.


In this article, we begin with a 2D topological insulator~(TI) and $d$-wave superconductor heterostructure to generate the 2D FSOTSC hosting $0$- and $\pi$-MCMs using a harmonic drive in the onsite mass term (see Fig.~\ref{static}). The signature of the MCMs is demonstrated in the local density of states~(LDOS) behavior (see Fig.~\ref{Harmonic}). Furthermore, the time-evolution of the MCMs is investigated where the $0$- and $\pi$-modes emerge even before the full time period  (see Fig.~\ref{TimeDynamics}). We topologically characterize the dynamical system using the average quadrupolar motion (see Fig.~\ref{Invariants}).  Afterward, we make use of the FPT to derive an effective Hamiltonian picture for the system in the strong driving amplitude limit (see Fig.~\ref{PeturbationEigenvalue}). 
We study the time dynamics of the $0$-MCMs for the exact and perturbative time-evolution operator and present a comparison between those two results in the strong driving amplitude limit (see Fig.~\ref{PeturbationBulkGapTimeDynamics}).


The remainder of the article is arranged as follows. We introduce our model Hamiltonian, the driving protocol, and the formalism in Sec.~\ref{Sec:II}. The numerical results obtained using the Floquet operator are discussed in Sec.~\ref{Sec:III}. We topologically characterize the dynamical $0$- and $\pi$-MCMs in Sec.~\ref{Sec:IV} with appropriate topological invariant. Sec.~\ref{Sec:V} is devoted to the discussion of analytical results obtained from the FPT and its comparison with the exact (numerical) results. In Sec.~\ref{Sec:VI}, we discuss some relevant points related to our results with their outlooks. Finally, we summarize and conclude our article in Sec.~\ref{Sec:VII}.

\section{Model and Method}\label{Sec:II}
In this section, we introduce our model Hamiltonian, driving protocol, and the formalism used to deal with the dynamical problem.

\subsection{Model Hamiltonian}
We consider a system consisting of a 2D TI in proximity to a $d$-wave superconductor~(see Fig.~\ref{static} for a schematic representation). The system can be described via the following Hamiltonian, written in the Bogoliubov-de Gennes (BdG) form as~\cite{Yan2018,ghosh2020floquet2}
\begin{eqnarray}\label{staticHam}
H(\vect{k})&=&2 \lambda \sin k_x \Gamma_1 + 2 \lambda \sin k_y \Gamma_2 + \epsilon (\vect{k}) \Gamma_3 \non \\
&&+ \Delta  \left(\cos k_x -\cos k_y\right) \Gamma_4 \ ,
\end{eqnarray}
where, $\epsilon(\vect{k})=(m_0 - 4 \gamma + 2 \gamma \cos k_x +2 \gamma \cos k_y )$, the hopping (SOC) strength is denoted by $\gamma~(\lambda)$; $m_0$ and $\Delta$ stand for the on-site crystal field splitting and $d$-wave superconducting pairing amplitude (assumed to be induced via the proximity effect), respectively. For static systems, such mean-field $d$-wave pairing 
has been theoretically considered before in Refs.~\cite{Yan2018,Liu2018} assuming proximity effect. The $8 \times 8$ $\vect{\Gamma}$ matrices are given by $\Gamma_1=\tau_z \sigma_z s_x,~\Gamma_2=\tau_z \sigma_0 s_y,~\Gamma_3=\tau_z \sigma_z s_0$, and $\Gamma_4=\tau_x \sigma_0 s_0$. The pauli matrices $\vect{\tau}$, $\vect{\sigma}$, and $\vect{s}$ operate on particle-hole~($e$, $h$), orbital~($\tilde{\alpha}, \tilde{\beta}$), and spin~($\ua, \da$) degrees of freedom, respectively. The Hamiltonian [\Eq{staticHam}] is invariant under time-reversal symmetry: $\mathcal{T} ^{-1} H(\vect{k}) \mathcal{T}=H(-\vect{k})$ and particle-hole symmetry: $\mathcal{C}^{-1} H(\vect{k}) \mathcal{C}=-H(-\vect{k})$, where $\mathcal{T}=i \tau_0 \sigma_0 s_y \mathcal{K}$ and $\mathcal{C}=\tau_y \sigma_0 s_y \mathcal{K}$ with $\mathcal{K}$ representing the complex-conjugation operator. Apart from these discrete symmetries, $H(\vect{k})$ also respect the following spatial symmetries: mirror symmetry along-$x$ with $\mathcal{M}_x= \tau_x \sigma_x s_0$: $\mathcal{M}_x H(k_x,k_y) \mathcal{M}_x^{-1}= H(-k_x,k_y) $, mirror symmetry along-$y$ with $\mathcal{M}_y= \tau_x \sigma_y s_0$: $\mathcal{M}_y H(k_x,k_y) \mathcal{M}_y^{-1}= H(k_x,-k_y) $, four-fold rotation with $C_4= \tau_z e^{-\frac{i \pi}{4}\sigma_z s_z}$: $C_4 H(k_x,k_y)C_4^{-1}= H(-k_y,k_x) $, mirror rotation-I with $\mathcal{M}_{xy}=C_4 \mathcal{M}_y$: $\mathcal{M}_{xy} H(k_x,k_y) \mathcal{M}_{xy}^{-1}= H(k_y,k_x)$, and mirror rotation-II with $\mathcal{M}_{x\bar{y}}=C_4 \mathcal{M}_x$: $\mathcal{M}_{x\bar{y}} H(k_x,k_y) \mathcal{M}_{x\bar{y}}^{-1}= H(-k_y,-k_x)$~\cite{Ghosh2022NHPRBL}. These symmetries play a pivotal role in the generation of FHOTSC, phase which we discuss later in Sec.~\ref{Sec:VIB}. The static Hamiltonian $H(\vect{k})$ manifests the second order TSC phase hosting two MZMs per corner when $0<  m_0  < 8 \gamma$ and $\Delta \neq 0$~\cite{Yan2018,ghosh2020floquet2}.

\subsection{Driving Protocol and Formalism}
We consider the following time-dependent harmonic drive~\cite{TitumPRB2017} (see Fig.~\ref{static} as a cartoon) in the on-site mass term on top of the static Hamiltonian $H(\vect{k})$~[\Eq{staticHam}] to realize the 2D FSOTSC as follows
\begin{eqnarray}\label{drive}
V(t)&=&  V \cos ( \Omega t)  \Gamma_3  \ ,
\end{eqnarray}
where, $V$ represents the strength of the drive and $\Omega=(2 \pi /T)$ is the frequency~(time-period) of the drive. It is evident that $V(t)$ satisfies $V(t+T)=V(t)$. Hence, the full Hamiltonian $\mathcal{H}(\vect{k},t)=H(\vect{k})+V(t)$ is also time-periodic \ie $\mathcal{H}(\vect{k},t+T)=\mathcal{H}(\vect{k},t)$. The Fourier components of $\mathcal{H}(\vect{k},t)$ reads as
\begin{equation}\label{FourierComp}
	\mathcal{H}_{\alpha} = \int_0^T \frac{d t}{T} \mathcal{H}(\vect{k},t) \ e^{i \alpha \Omega t} \ .
\end{equation}
Exploiting the frequency-zone scheme, one can construct the infinite dimensional time-independent Floquet Hamiltonian as~\cite{Eckardt2015}
\begin{widetext}
\begin{equation}\label{FloquetHam}
H^\infty_F(\vect{k})=
\begin{pmatrix}
\ddots& & &  \vdots & & & \udots\\
 &\mathcal{H}_{-2} & \mathcal{H}_{-1} & H (\vect{k}) -2\Omega &  \mathcal{H}_{1} &  \mathcal{H}_{2} \\
& & \mathcal{H}_{-2} & \mathcal{H}_{-1} & H (\vect{k}) -\Omega &  \mathcal{H}_{1} &  \mathcal{H}_{2} &    \\
& & & \mathcal{H}_{-2} & \mathcal{H}_{-1} &H (\vect{k}) &\mathcal{H}_{1} & \mathcal{H}_{2}\\
& & &  & \mathcal{H}_{-2} & \mathcal{H}_{-1} & H (\vect{k}) +\Omega &  \mathcal{H}_{1} &  \mathcal{H}_{2}  \\
& & & & & \mathcal{H}_{-2} & \mathcal{H}_{-1} & H (\vect{k}) +2\Omega &  \mathcal{H}_{1} &  \mathcal{H}_{2} \\
&&  & &  \udots & & &\vdots  & & \hspace{1cm}\ddots\\
\end{pmatrix} \ .
\end{equation}
\end{widetext}
However, the Fourier components ${\mathcal H}_\alpha$ having $\lvert \alpha \rvert > 1$ vanishes akin to the mathematical form of the drive. Here, the size of the Hamiltonian $H^\infty_F(\vect{k})$ acts as the bottleneck in the numerical studies, especially when either the frequency of the drive is comparable to (or less than) the bandwidth of the system or the drive encompasses higher-harmonics of the sinusoidal function in Eq.~(\ref{drive}). In such cases, we need to incorporate more Fourier components ${\mathcal H}_\alpha$, which in turn enlarge the size of $H^\infty_F(\vect{k})$. To circumvent this issue, we employ the time domain formalism and use the time-evolution operator $U(\vect{k};t,0)$ defined in terms of a time-ordered~(TO) notation as follows
\begin{eqnarray}\label{time-evolution}
U(\vect{k};t,0)&=& {\rm TO} ~ \exp \left[ -i \int_{0}^{t} dt' \mathcal{H}(\vect{k},t') \right]  \non \\
&=& \prod_{j=0}^{N-1} U(\vect{k};t_{j}+\delta t,t_j) \ ,
\end{eqnarray}
where, $U(\vect{k};t_{j}+\delta t,t_j)=e^{-i \mathcal{H}(\vect{k},t_j)\delta t}$; with $\delta t = \frac{t}{N}$ and $t_j=j \delta t$. However, $U(\vect{k};t_{j}+\delta t,t_j)$ can be calculated more efficiently using the second-order Trotter-Suzuki formalism as follows~\cite{DAlessio2015,Suzuki1976,DeRaedt183,QinPRB2022}
\begin{eqnarray}\label{TS_time-evo}
U(\vect{k};t_{j}+\delta t,t_j)= e^{- i  \frac{\delta t}{2} V\left(t_j + \frac{\delta t}{2} \right)}  e^{-i \delta t H(\vect{k})}  e^{- i  \frac{\delta t}{2} V\left(t_j + \frac{\delta t}{2} \right)}  . \ \ \
\end{eqnarray} 

Note that, we need to calculate $e^{-i \delta t H(\vect{k})}$ only once due to its time-independent nature. We choose the time increment $\delta t$ in such a way that $U(\vect{k};t,0)$ remains always unitary. Following \Eq{time-evolution}, we can construct the Floquet operator $U(\vect{k};T,0)$ by replacing $t \rightarrow T$, which in terms allows us to calculate the eigenvalue spectra and the LDOS therein. This $U(\vect{k};t,0)$ also facilitate the calculation of topological invariants, which we discuss in the forthcoming section. Following this periodic driving protocol [\Eq{drive}], 
one can generate the $0$- and $\pi$-MCMs depending upon the choice of the parameter space, which we present in the next section.

\section{Generation of anomalous MCM\lowercase{s} and their time dynamics}\label{Sec:III}
In this section, we provide all the numerical results obtained using the Floquet operator and the time-evolution operator discussed before.
\begin{figure}[]
	\centering
	\subfigure{\includegraphics[width=0.49\textwidth]{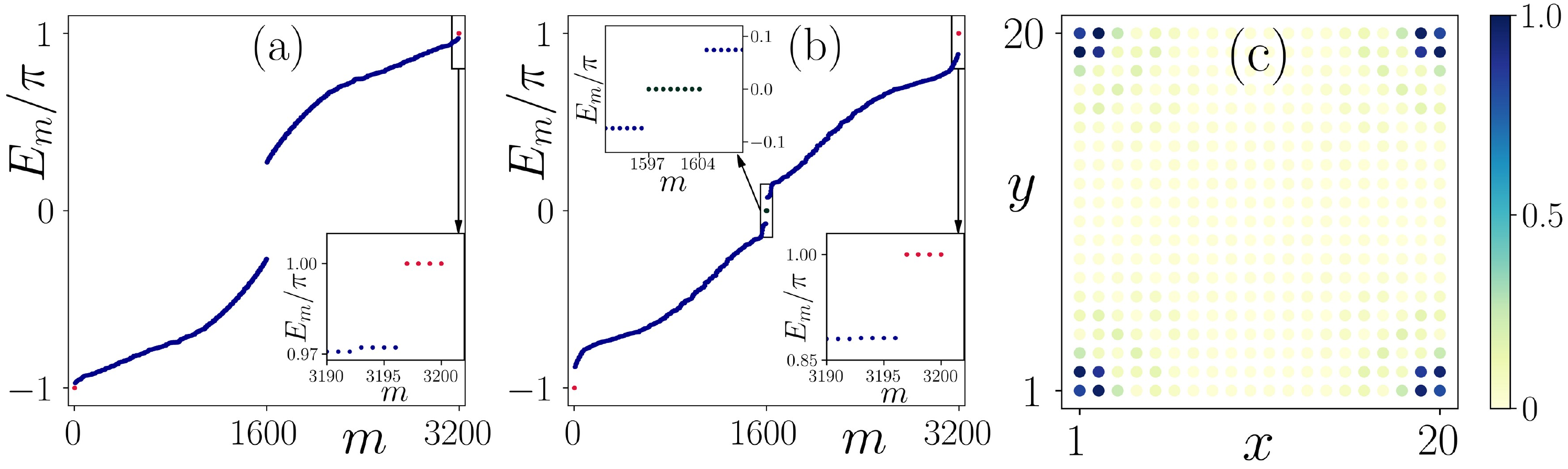}}
	\caption{We depict the quasi-energy spectra as function of the state index $m$ for the driven system starting from (a) non-topological regime ($m_0=-1.0$) and (b) topological regime ($m_0=1.0$). The $0$- and $\pi$-Majorana corner modes~(MCMs) are highlighted by dark-green and red dots, respectively. (c) The local density of states is illustrated for $0\mhyphen / \pi$ MCMs, which are sharply localized at the corners of the system. The other model parameters are chosen as $\gamma=\lambda=\Delta=0.2$, $V=1.0$, and frequency $\Omega=3.0$ and $1.5$ for panels (a) and (b), respectively.
	}
	\label{Harmonic}
\end{figure}
\subsection{Generation of anomalous MCMs}
As discussed earlier, we can obtain the Floquet operator $U(\vect{k};T,0)$ adopting \Eq{time-evolution}. We diagonalize the Floquet operator employing open boundary conditions~(OBCs) in both directions to acquire the eigenvalue spectra and signature of the localized boundary modes, \ie the MCMs therein. First, we consider that the system lies in the non-topological regime ($m_{0}=-1$) and then we harmonically drive the system to achieve the anomalous $\pi$-MCMs. We demonstrate the quasi-energy spectra as a function of the state index $m$ in Fig.~\ref{Harmonic}~(a), and the presence of the $\pi$-MCMs can be distinctly identified from the inset. The corresponding LDOS is shown in Fig.~\ref{Harmonic}~(c). Thus, it is evident that the $\pi$-MCMs are located at the corners of the system. However, one can also presume the static system to be in the topological regime ($m_{0}=1$), hosting the static MZMs, and the effect of the harmonic drive can be investigated there. We depict the quasi-energy spectra for one such scenario in Fig.~\ref{Harmonic}~(b). The driven system allows realization of both the regular $0$-MCMs and the anomalous $\pi$-MCMs, the latter does not have any static analog and can only be realized in a dynamical system. From the insets, one can identify the presence of both these modes. The LDOS distribution for the $0$- and $\pi$-MCMs are qualitatively the same as shown in Fig.~\ref{Harmonic}~(c). It is worth mentioning that the generation of the $0$- and $\pi$-MCMs are not limited to the specific choices of parameter sets and sustains for a considerable range in the parameter space. Nevertheless, we discuss the recipe to generate the FSOTSC in Sec.~\ref{Sec:VI}. 

\subsection{Time dynamics of the MCMs}
\begin{figure}[]
	\centering
	\subfigure{\includegraphics[width=0.49\textwidth]{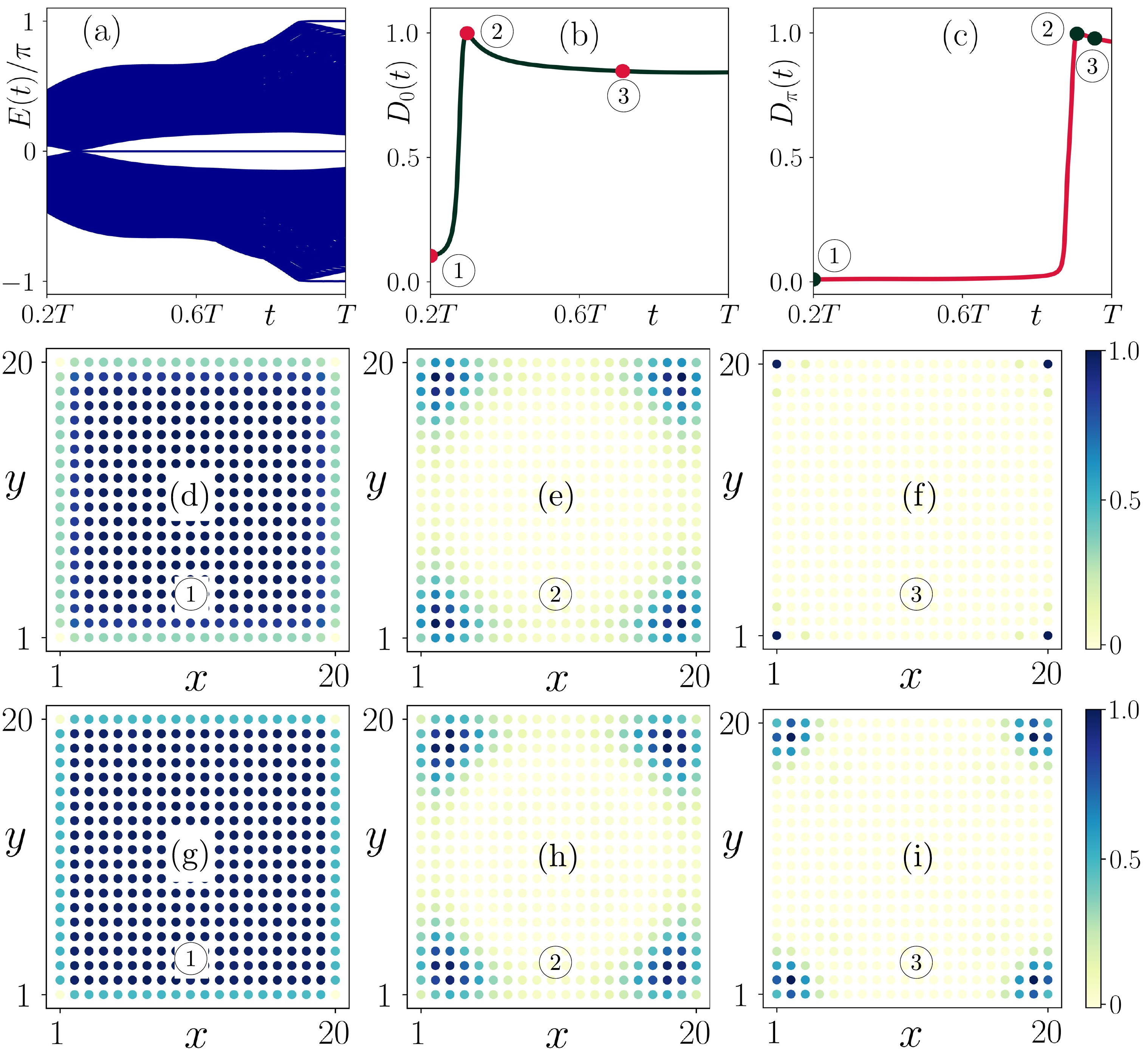}}
	\caption{(a) We demonstrate the time-dependent eigenvalue spectra $E(t)$ of the time-evolution operator $U(t,0)$ as a function of time $t$ during the driving period. The normalized total density 
of states at $E(t)=0$ and $E(t)=\pi$ are shown with respect to time $t$ in panels (b) and (c), respectively. We depict the local density of states (LDOS) at quasi-energy $0$ corresponding to the points marked in panel (b) as \textcircled{1} ($t=0.84$), \textcircled{2} ($t=1.25$), and \textcircled{3} ($t=3.0$), in panels (d), (e), and (f), respectively. In panels (f)-(i), we repeat (d)-(e) at quasi-energy $\pi$ corresponding to the designated points in panel (c) as \textcircled{1} ($t=0.84$), \textcircled{2} ($t=3.8$), and \textcircled{3} ($t=4.0$), respectively. Here, we choose the 
time-period $T=2\pi/\Omega\sim 4.2$ when $\Omega=1.5$.
	}
	\label{TimeDynamics}
\end{figure}

While exploring the time-evolution of the Floquet modes, we observe that the MCMs are not generated only after the full-time period $T$ rather they start appearing after a certain intermediate time. 
We can track the trails of the MCMs throughout the drive by analyzing the total density of states~(TDOS) of the MCMs present in the $0$- and $\pi$-gap. The TDOS at quasienergy $\epsilon$ can be computed as a function of $t$ as
\begin{equation}\label{TDOS}
D_\epsilon(t)= \sum_{m}  \delta\left[\epsilon-E_m(t)\right] \ .
\end{equation}
One can find $E_m(t)$ from the eigenvalues of $U(\vect{k};t,0)$. We illustrate $E(t)$ as a function of $t$ in Fig.~\ref{TimeDynamics}~(a). The appearance of in-gap $0$- and $\pi$-modes can be observed with respect to time from the eigenvalue spectra. The time-evolution highlights that both the $0$- and $\pi$-modes emerge when $t<T$ and remain there at $t=T$. However, for better identification, we show the time dependent TDOS corresponding to quasi-energy $0$ and $\pi$ in Fig.~\ref{TimeDynamics}~(b) and (c), respectively. By analyzing the Fig.~\ref{TimeDynamics}~(b) and (c), we can reckon that both the regular $0$- and $\pi$-MCMs emerge at a time $t<T$ and prevails throughout the driving period $T$. One can also investigate the localization properties of the $0$- 
and $\pi$-MCMs, specifically when they start to appear. We choose three time-points associated with Fig.~\ref{TimeDynamics}(b)~[(c)]: \circled{1} with no $0$-MCMs~[$\pi$-MCMs], \circled{2} the transition point, when $0$-MCMs~[$\pi$-MCMs] start to appear, and \circled{3} after generation of the $0$-MCMs~[$\pi$-MCMs]. We depict the LDOS at quasi-energy $0$~[$\pi$] corresponding to the points \circled{1}, \circled{2}, and \circled{3} in Fig.~\ref{TimeDynamics} (d), (e), and (f), respectively~[Fig.~\ref{TimeDynamics} (g), (h), and (i), respectively]. From Fig.~\ref{TimeDynamics}~(d) and (g), 
one can observe that no boundary modes exists when $t\ll T$ \ie~no $0$-/$\pi$-MCMs are generated. At the transition point (see Fig.~\ref{TimeDynamics}~(e) and (h)), the $0$-/$\pi$-modes start to populate near the corners of the system. However, we observe sharp localization near the corners of the system for $0$- and $\pi$-modes in Fig.~\ref{TimeDynamics} (f) and (g), respectively. Therefore, one can infer a connection between the time evolution of the MCMs (shown via the total density of states) to the eigenvalues of the time evolution operator at time $t$ as depicted via the LDOS.

\section{Topological Characterization of the MCM\lowercase{s}}\label{Sec:IV}
\begin{figure}[]
	\centering
	\subfigure{\includegraphics[width=0.49\textwidth]{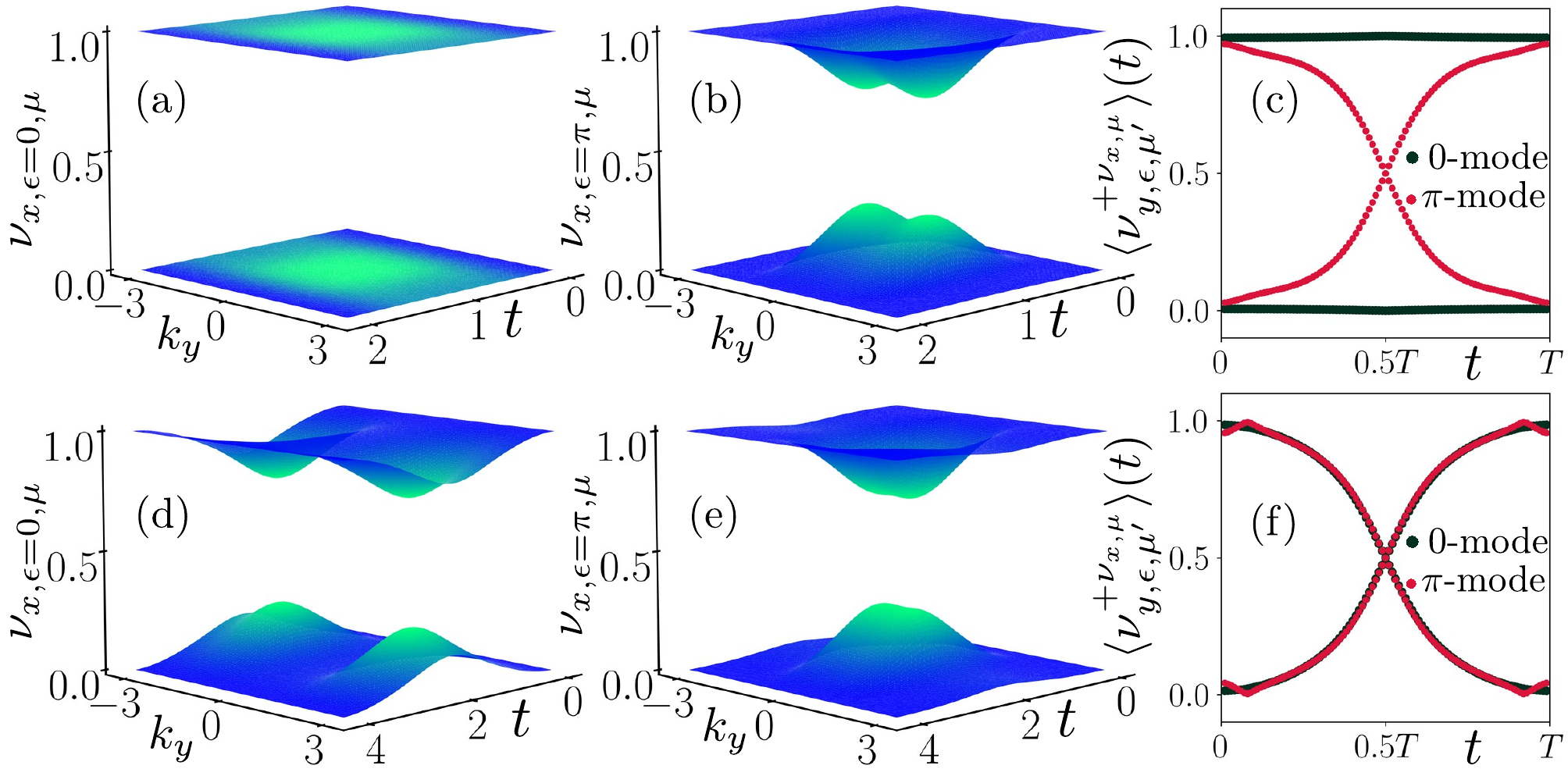}}
	\caption{We show the dynamical first-order branches $\nu_{x,\epsilon}$ in the $k_y \mhyphen t$ plane for $0$-gap in panel (a) and for $\pi$-gap in panel (b), corresponding to Fig.~\ref{Harmonic}~(a).  The resulting average quadrupolar motion $\langle \nu_{y,\epsilon , \mu'}^{+ \nu_{x,\epsilon}} \rangle (t)$ is depicted as a function of $t$ in panel (c). The first-order branches are always gapped during the time interval $t\in\left[0,T\right]$. Since, only $\pi$-Majorana corner modes~(MCMs) are present in this case, hence $\langle \nu_{y,\epsilon , \mu'}^{+ \nu_{x,\epsilon}} \rangle (t)$ crosses the $0.5$ line for the $\pi$-gap only. We show the dynamical first-order branches and the average quadrupolar motion, similar to panels (a)-(c), corresponding to the case of Fig.~\ref{Harmonic}~(b), in panels (d)-(f). Since, both the $0$- and $\pi$-MCMs are present in this case, $\langle \nu_{y,\epsilon , \mu'}^{+ \nu_{x,\epsilon}} \rangle (t)$ crosses the $0.5$ line for both the gaps.
	}
	\label{Invariants}
\end{figure}
We topologically characterize the regular $0$- and anomalous $\pi$-MCMs employing the \textit{dynamical nested Wilson loop technique}. The detailed discussions regarding this method can be found in Refs.~\cite{HuPRL2020,GhoshDynamical2022}. Thus, we do not repeat the same here. To construct the dynamical second-order nested Wilson loop operator we need the notion of a gap in which the MCMs appear. We incorporate the same via the periodized evolution operator, defined as~\cite{Rudner2013,HuPRL2020,GhoshDynamical2022} 
\begin{eqnarray}\label{periodizedoperator}
U_{\epsilon}(\vect{k};t,0)=U(\vect{k};t,0)\left[U(\vect{k};T,0)\right]^{-t/T}_\epsilon \ .
\end{eqnarray}
Here, the subscript $\epsilon$ represents the $0$- and $\pi$-gap. We can construct $U_{\epsilon}(\vect{k};t,0)$ in a straight forward manner using \Eq{time-evolution}. Afterward, we construct the dynamical first-order Wilson loop operator for the $\epsilon$-gap as~\cite{HuPRL2020,GhoshDynamical2022}
\begin{eqnarray}
W_{x,\epsilon,\vect{k}}(t)= Q_{x,\epsilon,\vect{k}+(L_x-1)\Delta_x \vect{e}_x}(t) \cdots Q_{x,\epsilon,\vect{k}+\Delta_x \vect{e}_x}(t)  Q_{x,\epsilon,\vect{k}}(t) , \non \\
\end{eqnarray}
where, $Q_{x,\epsilon,\vect{k}}(t)=\frac{\mathbb{I} + U^\dagger_{\epsilon}(\vect{k}+\Delta_x\vect{e}_x;t,0) U_{\epsilon}(\vect{k};t,0)}{2}$; with $\Delta_x=2 \pi / L_x$ and $\vect{e}_x$ representing unit vector along the $x$-direction. The eigenvalue equation for $W_{x,\epsilon,\vect{k}}(t)$ can be written in the form
\begin{eqnarray}
W_{x,\epsilon,\vect{k}}(t) \ket{\nu_{x,\epsilon , \mu}(\vect{k},t)} = e^{-2 \pi i \nu_{x,\epsilon, \mu}(k_y,t)}\ket{\nu_{x,\epsilon , \mu}(\vect{k},t)} \ , \quad \quad
\end{eqnarray}
Here, $\nu_{x,\epsilon, \mu}(k_{y},t)$ represents the dynamical first-order branches. We illustrate $\nu_{x,\epsilon, \mu}(k_{y},t)$ in Figs.~\ref{Invariants}~(a), (b), and (d), (e). In the former case, the system exhibits only $\pi$-MCMs, whereas, the later case supports both the $0$- and $\pi$-MCMs. Interestingly, in both the cases $\nu_{x,\epsilon, \mu}(k_{y},t)$ exhibit a gap throughout the driving-period $T$ and the variation of crystal momentum $k_y$. It therefore yields us an ideal platform to construct the dynamical second-order (or nested) Wilson loop operator~\cite{HuPRL2020,GhoshDynamical2022}. 
Owing to the gapped nature of the $\nu_{x,\epsilon, \mu}(k_{y},t)$, one can configure two sets out of the eight $\nu_{x,\epsilon}$ such that $\nu_{x,\epsilon,\mu}\in \pm \nu_{x,\epsilon}$. We construct the dynamical nested Wilson loop operator in the subspace of $\pm \nu_{x,\epsilon}$ as

\begin{widetext}
\begin{eqnarray}
W_{y,\epsilon,\vect{k}}^{\pm \nu_{x,\epsilon}} (t)= Q_{y,\epsilon,\vect{k}+(L_y-1)\Delta_y \vect{e}_y}^{\pm \nu_{x,\epsilon}}(t) \ \ \cdots  \ \ Q_{y,\epsilon,\vect{k}+\Delta_y \vect{e}_y}^{\pm \nu_{x,\epsilon}}(t) \ Q_{y,\epsilon,\vect{k}}^{\pm \nu_{x,\epsilon}}(t) \ ,
\end{eqnarray}
with $\left[Q_{y,\epsilon,\vect{k}}^{\pm \nu_{x, \epsilon}} (t)\right]_{\mu_1 \mu_2} = \sum_{m n} \left[\nu_{x,\epsilon , \mu_1} (\vect{k}+\Delta_y \vect{e}_y,t) \right]^*_m $ $ \left[Q_{y,\epsilon,\vect{k}}(t)\right]_{mn} \left[\nu_{x,\epsilon , \mu_2} (\vect{k},t) \right]_n$ ; where,  $Q_{y,\epsilon,\vect{k}}(t)=\frac{\mathbb{I} + U^\dagger_{\epsilon}(\vect{k}+\Delta_y\vect{e}_y;t,0) U_{\epsilon}(\vect{k};t,0)}{2}$, $\Delta_y=2 \pi /L_y$, and $\vect{e}_y$ is the unit vector along the $y$-direction. We can write down the eigenvalue equation for $W_{y,\epsilon,\vect{k}}^{\pm \nu_{x,\epsilon}} (t)$ as
\begin{eqnarray}
W_{y,\epsilon,\vect{k}}^{\pm \nu_{x,\epsilon}} (t) \ket{\nu_{y,\epsilon , \mu'}^{\pm \nu_{x,\epsilon}}(\vect{k},t)} = e^{-2 \pi i \nu_{y,\epsilon, \mu'}^{\pm \nu_{x,\epsilon}}(k_{x},t)}\ket{\nu_{y,\epsilon , \mu'}^{\pm \nu_{x,\epsilon}}(\vect{k},t)} \ , 
\end{eqnarray}
\end{widetext}
where, $\nu_{y,\epsilon, \mu'}^{\pm \nu_{x,\epsilon}}(k_{x},t)$ represents the dynamical second-order branches. Here, the $k_x$ dependence of the $\nu_{y,\epsilon, \mu'}^{\pm \nu_{x,\epsilon}}(k_{x},t)$ originates from the fact that the eigenfunctions of the dynamical first-order Wilson loop operator depends explicitly on the choice of the base point. However, its eigenvalues are independent of the specific choice of the base point. We define the average quadrupolar motion as~\cite{HuPRL2020,GhoshDynamical2022} 
\begin{eqnarray}\label{quadrupolarmotion}
\langle \nu_{y,\epsilon , \mu'}^{+ \nu_{x,\epsilon}} \rangle (t)=\frac{1}{L_x} \sum_{k_x} \nu_{y,\epsilon , \mu'}^{+ \nu_{x,\epsilon}} (k_x,t) \ .
\end{eqnarray}
Note that, the particles complete a round trip during the time-interval $t \in \left[0,T\right]$, since $U_{\epsilon}(\vect{k};0,0)=U_{\epsilon}(\vect{k};T,0)=\mathbb{I}$. We hence obtain two fixed-points: $\langle \nu_{y,\epsilon , \mu'}^{+ \nu_{x,\epsilon}} \rangle (t=0)=\langle \nu_{y,\epsilon , \mu'}^{+ \nu_{x,\epsilon}} \rangle (t=\frac{T}{2})=0~({\rm mod}~1)$. For a topologically trivial system, $\langle \nu_{y,\epsilon , \mu'}^{+ \nu_{x,\epsilon}} \rangle (t)$ starts from $0~({\rm mod}~1)$ and evolves back to $0~({\rm mod}~1)$ during $t \in \left[0,T\right]$ and their motion can be adiabatically connected to zero, when boundary is imposed~[see Fig.~\ref{Invariants}~(c) for $0$-gap]. However, if there is any obstruction due to the presence of in-gap states, \ie the MCMs, the quadrupolar motion $\langle \nu_{y,\epsilon , \mu'}^{+ \nu_{x,\epsilon}} \rangle (t\rightarrow 0) = 0~(1) $ ends up at $\langle \nu_{y,\epsilon , \mu'}^{+ \nu_{x,\epsilon}} \rangle (t\rightarrow T)= 1~(0)$ and the two branches cross each other 
at $0.5~({\rm mod}~1)$ [see Fig.~\ref{Invariants}~(c) for $\pi$-gap and (f) for both $0$- and $\pi$-gap]. Such kind of motion cannot be connected to zero and thus can be utilized to distinguish between the topological and non-topological phases, in the dynamical case~\cite{HuPRL2020,GhoshDynamical2022}.

\section{Analytical approach: Floquet Perturbation Theory}\label{Sec:V}

\begin{figure}[]
	\centering
	\subfigure{\includegraphics[width=0.49\textwidth]{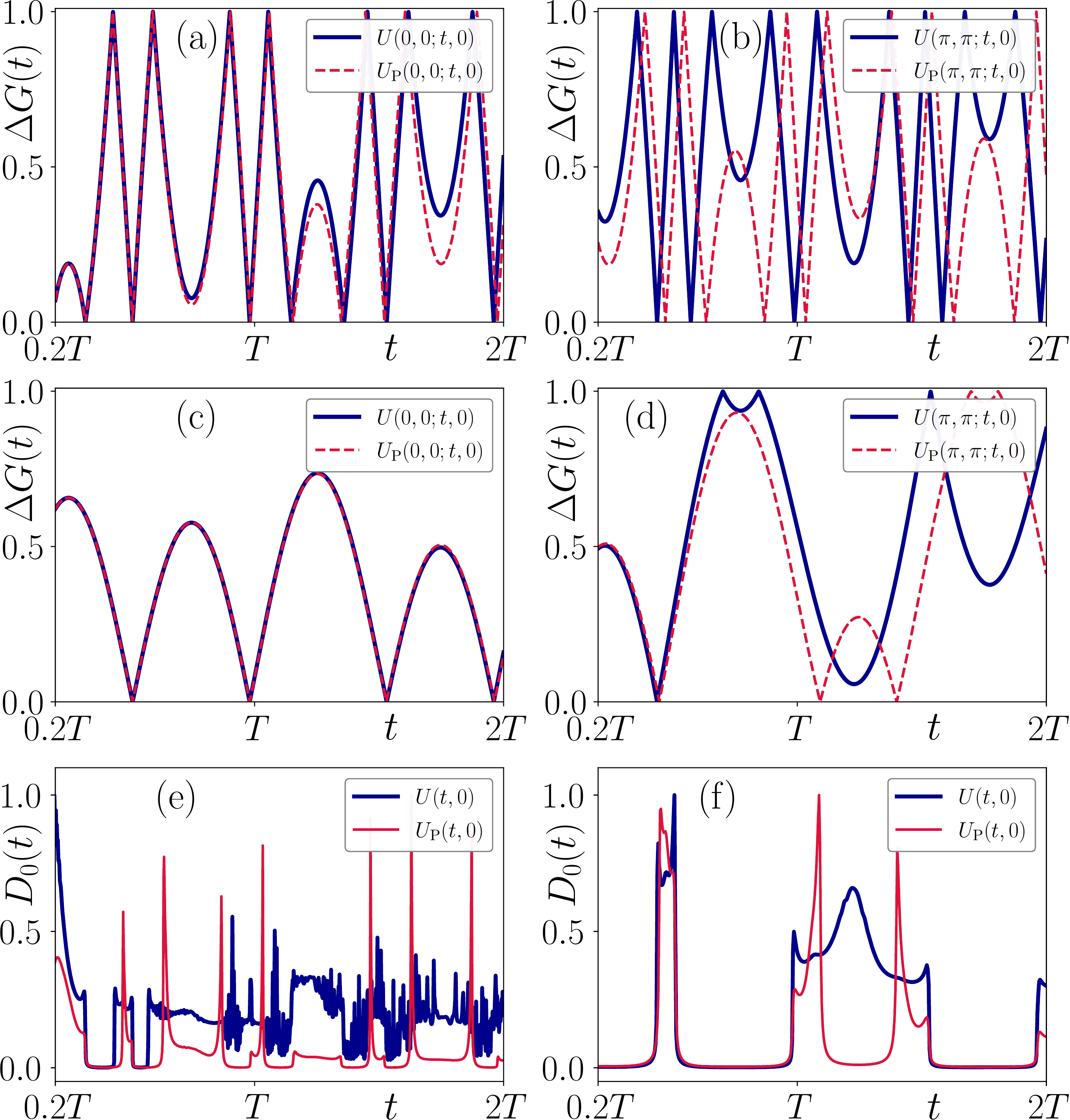}}
	\caption{We depict the bulk gap $\Delta G(t)$ as a function of $t$ at $\vect{k}=(0,0)$ obtained from the exact Floquet operator $U(\vect{k};t,0)$~(solid blue line) and the perturbative Floquet operator $U_{\rm P}(\vect{k};t,0)$~(dashed red line) for (a) $\Omega=1.5$ and (c) $\Omega=5.0$. We show the same at $\vect{k}=(\pi,\pi)$ in panels (b) and (d) respectively. We illustrate the total density of states $D_0(t)$ as function of $t$ corresponding to $\Omega=1.5$ and $\Omega=5.0$ in panels (e) and (f), respectively. Here, blue and red lines represent the total density of states $D_0(t)$ computed from 
	$U(t,0)$ and $U_{\rm P}(t,0)$, respectively. We choose $V=10.0$ and $m_0=0.2$, while all other model parameters take the same value as mentioned in Fig.~\ref{Harmonic}.
	}
	\label{PeturbationBulkGapTimeDynamics}
\end{figure}

After investigating our setup numerically, we tie-up with the FPT~\cite{Sen2021,Mukherjee2020,Sengupta2022Weyl} to achieve some analytical insights into the problem. Within the analytic scheme, we assume the amplitude of the drive $V$ to be much larger than the hopping amplitude $\gamma$. We treat $V(t)$ [\Eq{drive}] exactly and $H(\vect{k})$ [\Eq{staticHam}] as perturbation. The time-evolution operator 
for $V(t)$ reads as
\begin{eqnarray}
U_0(t,0)&=&\exp \left[ -i \int_{0}^{t} dt' V(t') \right] \non \\
&=&\exp \left[  -\frac{i V}{\Omega} \sin (\Omega t) \ \Gamma_3  \right]\ .
\end{eqnarray}
We employ the interaction picture to obtain the full time-evolution operator. Following the perturbation theory~\cite{Sen2021,Mukherjee2020,Sengupta2022Weyl}, the latter can be written in the form
\begin{eqnarray}\label{Upert}
U_{\rm P}(\vect{k};t,0)= U_0(t,0) \ U_{\rm I} (\vect{k};t,0) \ ,
\end{eqnarray}
where, the time-evolution operator in the interaction picture $U_{\rm I} (\vect{k};t,0)$ can be written in the form of a power-series as~\cite{Sen2021,Mukherjee2020,Sengupta2022Weyl}
\begin{widetext}
\begin{eqnarray}\label{pertseries}
	U_{\rm I}(\vect{k};t,0)&=& \mathbb{I} + (-i) \int_{0}^{t} dt' H_{\rm I}(\vect{k},t') + (-i)^2 \int_{0}^{t} dt_1 H_{\rm I}(\vect{k},t_1)  \int_{0}^{t_1} dt_2 H_{\rm I}(\vect{k},t_2) + \cdots \non \\
	&=& \mathbb{I} + U_{\rm I}^{(1)}(\vect{k};t,0) +  U_{\rm I}^{(2)}(\vect{k};t,0) + \cdots  \ ,
\end{eqnarray}
where, $H_{\rm I}(\vect{k},t)=U_0(0,t) H(\vect{k}) U_0(t,0)$. 
Within our analysis, we truncate the series after the first-order term in $U_{\rm I}(\vect{k};t,0)$ as the higher order terms become messy large expressions with smaller in magnitude. Following the periodic cosine drive [\Eq{drive}], we obtain the first-order term $U_{\rm I} (t,0)$ in the perturbation series [\Eq{pertseries}] as 
\begin{eqnarray}\label{firstorderperturbation}
	U_{\rm I}^{(1)} (\vect{k};t,0) &=&-\frac{i}{2} \Bigg[ \left\{ t + t \mathcal{J}_0 (\phi) + 2 \sum_{n=1}^{\infty} \mathcal{J}_{2n} (\phi) \frac{\sin 2n \Omega t }{2 n \Omega}\right\} H(\vect{k}) +2 i \sum_{n=1}^{\infty} \frac{ \mathcal{J}_{2n-1}(\phi)} {(2n-1) \Omega}  \left\{ 1 - \cos \left[ (2n-1) \Omega t \right] \right\} \left[\Gamma_3, H(\vect{k}) \right] \non \\ 
	&& + \left\{ t - t \mathcal{J}_0 (\phi) + 2 \sum_{n=1}^{\infty} \mathcal{J}_{2n} (\phi) \frac{\sin 2n \Omega t }{2 n \Omega}\right\} \Gamma_3 H(\vect{k})  \Gamma_3 \Bigg] \ ,
\end{eqnarray}
\end{widetext}
where, $\phi=\frac{2 V}{\Omega}$ and $\mathcal{J}_n$ is the Bessel function of first kind. However, $U_{\rm P}(\vect{k};t,0)$ does not satisfy the unitarity condition. It is possible to recast the same at 
$t=T$ by exponentiating $U_{\rm I}(\vect{k};t,0)$ and exploiting $U_0(t,0)=\mathbb{I}$. Due to the loss of unitarity condition, $U_{\rm P}(\vect{k};t,0)$ does not facilitate  the calculation of average quadrupolar motion~[\Eq{quadrupolarmotion}]. 
Following the first-order term in $U_{\rm I}(\vect{k};t,0)$~[\Eq{firstorderperturbation}] and combining this with $U_0(t,0)$, the effective Hamiltonian can be computed at $t=T$ from $U_{\rm P}(\vect{k};T,0)$~[\Eq{Upert}]. 
The effective Hamiltonian can be obtained from the relation $U_{\rm P}(\vect{k};T,0)=\exp \left[-i H_{\rm eff}(\vect{k})T\right]$ as
\begin{eqnarray}\label{EffHam}
	H_{\rm eff}(\vect{k})= \mathcal{J}_0 (\phi) \left[H(\vect{k})- \epsilon(\vect{k}) \right]  +  \epsilon(\vect{k}) \Gamma_3 \  . 
\end{eqnarray}
It is evident from the $H_{\rm eff}(\vect{k})$ that only the SOC and the $d$-wave pairing term get modulated by the drive, whereas the hopping, as well as the crystal field splitting term $m_0$, is unaffected by the application of the drive.  The bulk gap $\Delta G (T)$, obtained from $H_{\rm eff}(\vect{k})$,  at $\vect{k}=(0,0)$ and $(\pi,\pi)$ remain unaltered as compared to the static band gap obtained from  $H(\vect{k})$. However, the instantaneous bulk gap $\Delta G (t)$ as function of time $t (\neq T)$ can be modulated with respect to the static band gap due to the presence of other 
time-dependent non-vanishing terms in \Eq{firstorderperturbation} [see Figs.~\ref{PeturbationBulkGapTimeDynamics}~(a)-(d)].

\begin{figure}[]
	\centering
	\subfigure{\includegraphics[width=0.49\textwidth]{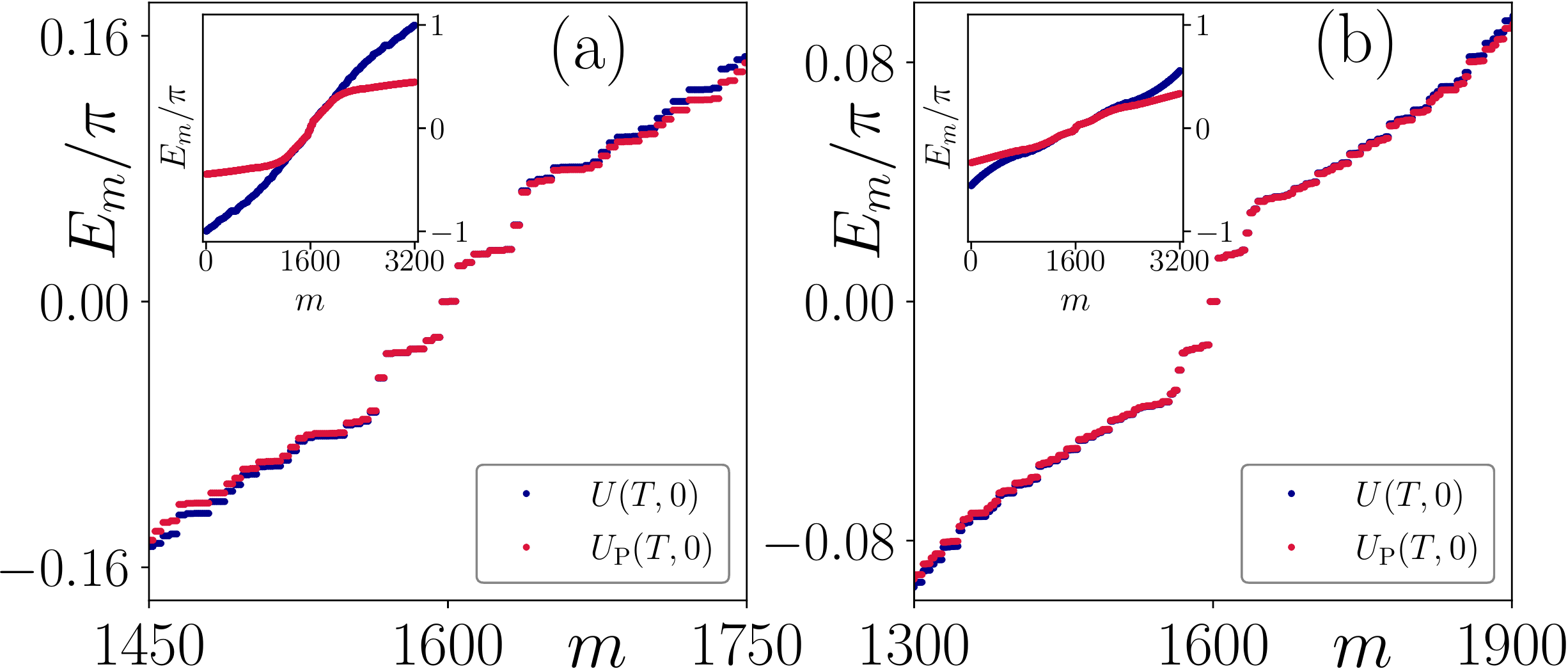}}
	\caption{We depict the quasi-energy spectra (employing open boundary condition) corresponding to the exact Floquet operator $U(T,0)$~(blue dots) and compare it with that of the perturbative Floquet operator $U_{\rm P}(T,0)$~(red dots) for the few low lying states in panel (a) for $\Omega=1.5$ and panel (b) for $\Omega=5.0$. We show the corresponding full quasi-energy spectra in the 
insets of the figures. We choose $V=10.0$ and $m_0=0.2$, while all the other model parameters are chosen to be of the same value as mentioned in Fig.~\ref{Harmonic}.
	}
	\label{PeturbationEigenvalue}
\end{figure}

In Fig.~\ref{PeturbationEigenvalue}, we depict the quasi-energy spectra (under OBC) for the Floquet operator constructed using \Eq{Upert} by substituting $t=T$. This is represented by red dots in Fig.~\ref{PeturbationEigenvalue}. We further compare the same with that of the exact Floquet operator $U(T,0)$~(indicated by blue dots in Fig.~\ref{PeturbationEigenvalue}). For a fixed value of 
$V~(\gg\gamma)$, we consider two cases for the driving frequency $\Omega=1.5$ and $\Omega=5.0$ as demonstrated in Fig.~\ref{PeturbationEigenvalue} (a) and (b), respectively. The quasi-energy spectra of the $U_{\rm P}(T,0)$ (obtained from FPT) match well with the low-lying eigenvalues $E_m \to 0$ of the exact Floquet operator $U(T,0)$. 
However, from the insets of Fig.~\ref{PeturbationEigenvalue}~(a) and (b), one can notice that the full eigenvalue spectra of $U_{\rm P}(T,0)$ do not fully match with that of $U(T,0)$. Nevertheless, it is evident from Fig.~\ref{PeturbationEigenvalue} (b) that the overlap between the exact quasi-energy (numerical) and the perturbative quasi-energy (analytical) becomes more prominent as $\Omega=5.0$. The same is not true in Fig.~\ref{PeturbationEigenvalue} (a) as $\Omega=1.5$ is comparable with the bandwidth. Interestingly, $U_{\rm P}(T,0)$ can capture the $0$-MCMs accurately in its eigenvalue spectra. However, $U_{\rm P}(T,0)$ fails to encapsulate the anomalous $\pi$-modes.

Having investigated the quasi-energy spectra, obtained employing OBC and depicted in Fig.~\ref{PeturbationEigenvalue}, we tend towards the time-dynamics of the MCMs within the FPT scheme. 
First, we compare the bulk gap $\Delta G(t)$ corresponding to the quasi-energy bands as a function of $t$ around the momentum $\vect{k}=(0,0)$ and $\vect{k}=(\pi,\pi)$, evaluated from $U_{\rm P}(\vect{k};t,0)$ and $U(\vect{k};t,0)$, in Fig.~\ref{PeturbationBulkGapTimeDynamics}~(a) and (b), respectively for $\Omega=1.5$. The same has been depicted in Fig.~\ref{PeturbationBulkGapTimeDynamics}~(c) and (d) choosing high frequency $\Omega=5.0$. The bulk-gap closing points (in the time axis) at $\vect{k}=(0,0)$ obtained from $U_{\rm P}(0,0;t,0)$ qualitatively coincides with that of $U(0,0;t,0)$~[see Fig.~\ref{PeturbationBulkGapTimeDynamics}~(a) and (c)]. However, at $\vect{k}=(\pi,\pi)$, we notice that there is a little mismatch between the gap closing points (in the time axis) obtained from $U_{\rm P}(\pi,\pi;t,0)$ and $U(\pi,\pi;t,0)$ \ie the analytical and the numerical approach. At higher frequency, the agreement between these two become more prominent up to the time-period $T$ as shown in Fig.~\ref{PeturbationBulkGapTimeDynamics}~(b) and (d). Although, as we increase the time above the time-period $T$ ($t>T$), we observe that there is a clear mismatch between the numerical and the analytical approach even for $\Omega=5.0$~[see Fig.~\ref{PeturbationBulkGapTimeDynamics}~(d)]. This mismatch remains as we increase the frequency 
$\Omega=8.0$ (not shown). We believe that this mismatch occurs due to the truncation of the evolution operator $U_{\rm I} (t,0)$ to the first-order term in the perturbation series [Eq.~(\ref{pertseries})].
By investigating the bulk-gap closing between two time-intervals, we can predict the apperance or disapperance of the MCMs in the finite-size system at that said interval. 

Nevertheless, we obtain a more clear picture of the time-evolution of the MCMs from the investigation of the TDOS $D_0(t)$~[\Eq{TDOS}]. We show $D_0(t)$ as a function of time $t$ for a finite-size system in Fig.~\ref{PeturbationBulkGapTimeDynamics}~(e) and (f) for $\Omega=1.5$ and $\Omega=5.0$, respectively. The blue and red lines represent $D_0(t)$ calculated from $U(t,0)$ and $U_{\rm P}(t,0)$, respectively. For $\Omega=1.5$, $D_0(t)$ calculated from $U_{\rm P}(t,0)$ overlaps with that of $U(t,0)$ only for a limited range of time. Nonetheless, both the exact and pertubative calculation suggest that the MCMs survives not only at $t=T$, but at a later time also. In case of $\Omega=5.0$, the $D_0(t)$ calculated using exact and perturbative Floquet operator match substantially over the entire time window $t\in [0,2T]$. The small mismatch accounts for the nonoverlapping of their bulk gap closing points between $U(\pi,\pi;t,0)$ and $U_{\rm P}(\pi,\pi;t,0)$. Another important point to 
note, for $\Omega=1.5$, the frequent fluctuations in the TDOS~[see Fig.~\ref{PeturbationBulkGapTimeDynamics}~(e)] appear since the instantaneous bulk gap $\Delta G(t)$ vanishes many times in the same interval~[see Fig.~\ref{PeturbationBulkGapTimeDynamics}~(a) and (b)]. However, for $\Omega=5.0$, the number of bulk gap vanishing points (in the time axis) is less [see Fig.~\ref{PeturbationBulkGapTimeDynamics}~(c) and (d)]. Hence, the TDOS obtained for this case is smoother in nature as compared to $\Omega=1.5$~[see Fig.~\ref{PeturbationBulkGapTimeDynamics}~(f)].

\section{Discussions and Outlook}\label{Sec:VI}
\begin{figure}[]
	\centering
	\subfigure{\includegraphics[width=0.49\textwidth]{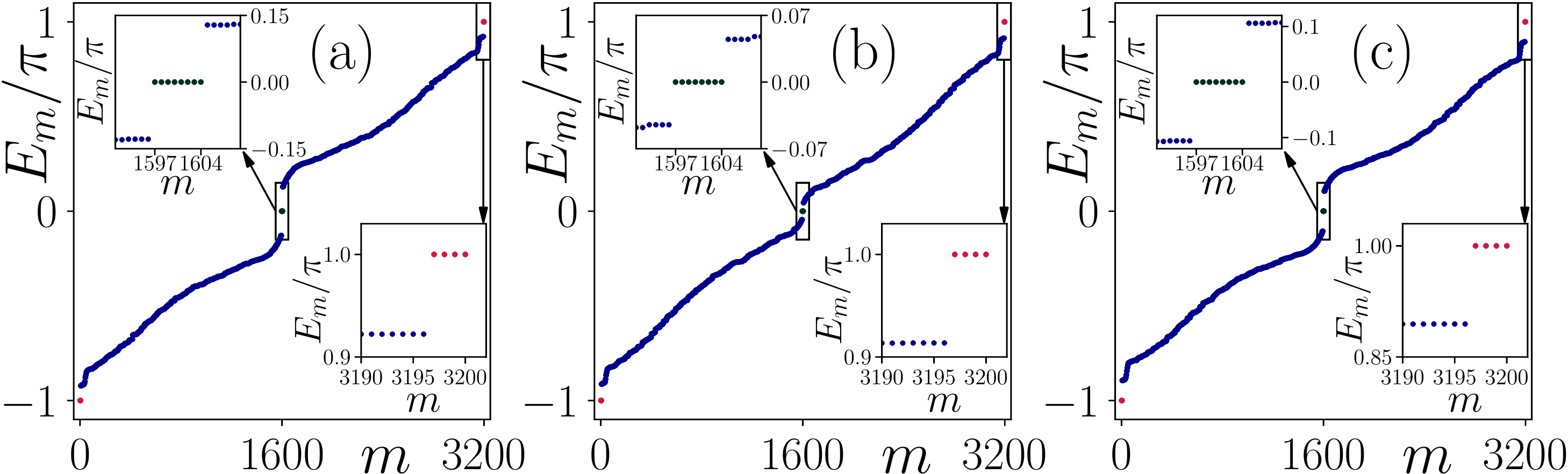}}
	\caption{In panels (a), (b), and (c), we illustrate the quasi-energy spectra as a function of the state index $m$ following the drive protocols as introduced in Eqs.~(\ref{drive2a}), (\ref{drive2b}), 
	and (\ref{drive2c}), respectively. We choose $\Omega=1.5$, while all other model parameters take the same value as mentioned in Fig.~\ref{Harmonic}.
	}
	\label{OtherDrivingProtocols}
\end{figure}
In this section, we discuss various aspects of the harmonic drive protocol and compare our FPT with the Brillouin-Wigner~(BW) perturbation theory~\cite{MikamiBW2016} that is valid in the 
high-frequency regime.

\subsection{Different driving protocols}
The formalism we introduced in Sec.~\ref{Sec:II} is not limited to the specific form of the drive [Eq.~(\ref{drive})]. One can introduce more sinusoidal functions in Eq.~(\ref{drive}) and study the effect of 
such drives. In this regard, we propose the following driving protocols:
\begin{subequations}
	\begin{eqnarray}
	V_1(t)&=& V \sin ( \Omega t)  \Gamma_3  \ , \label{drive2a}  \\
	V_2(t)&=& V \left[ \cos ( \Omega t) +\sin ( \Omega t) \right] \Gamma_3  \ , \label{drive2b} \\
	V_3(t)&=& V \left[ \cos ( \Omega t) + \cos ( 2 \Omega t) + \cos (3 \Omega t)\right] \Gamma_3  \label{drive2c} \ . \quad 
	\end{eqnarray}
\end{subequations}
Following the drive protocols mentioned in \Eq{drive2a}, (\ref{drive2b}), and (\ref{drive2c}), we show the quasi-energy spectra for a finite-size system in Fig.~\ref{OtherDrivingProtocols}~(a), (b), and (c), respectively. In all the three cases, the qualitative behavior of the quasi-energy spectra appear to be the similar as discussed before. The regular $0$- and anomalous $\pi$-modes appear in all such 
drive protocols.

\subsection{Recipe for generating the FSOTSC}\label{Sec:VIB}
\begin{figure}[]
	\centering
	\subfigure{\includegraphics[width=0.49\textwidth]{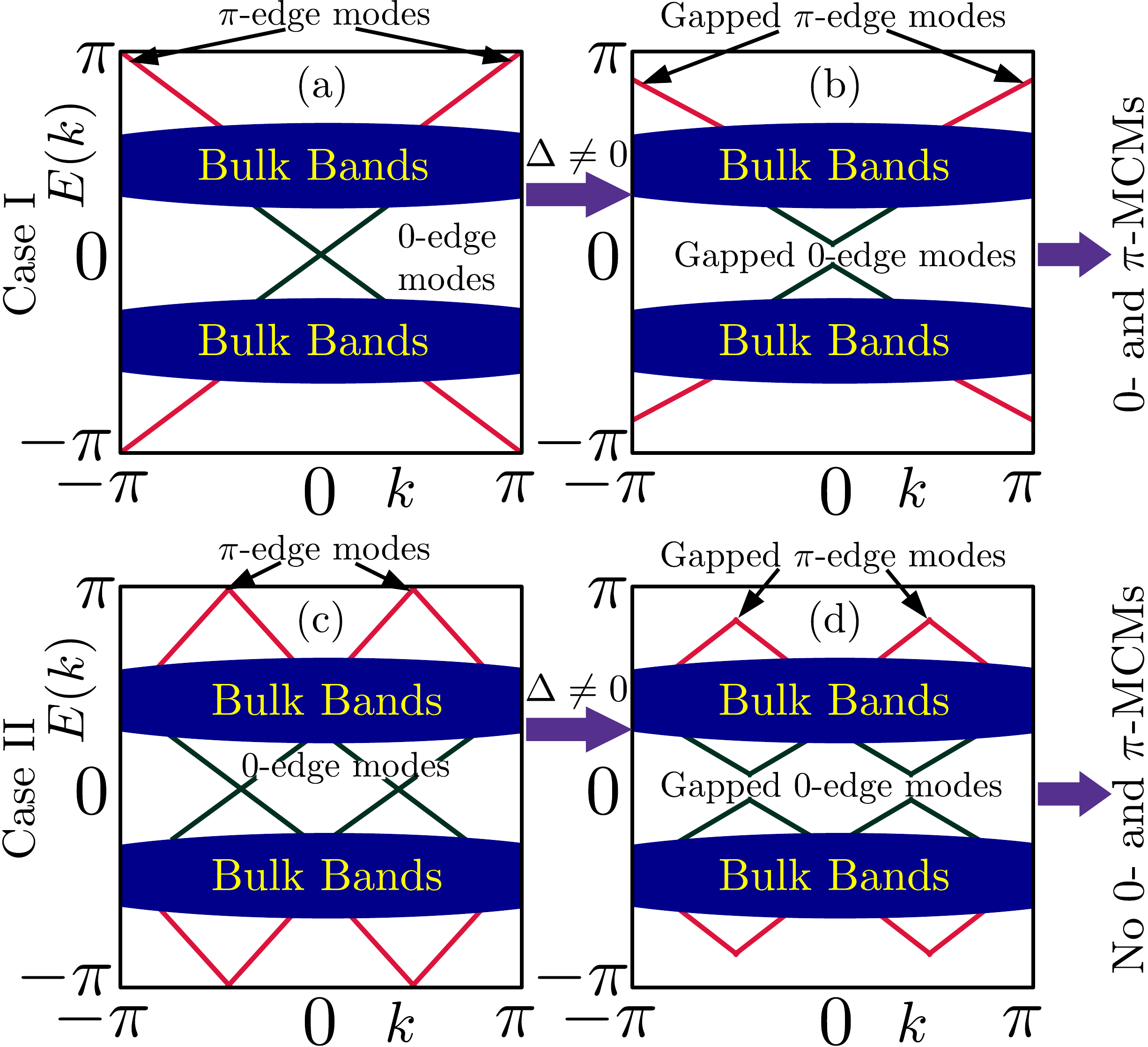}}
	\caption{We depict the schematic representation of the bulk bands and the edge modes, at $k=0, \pm \pi$ (case I) and  $k \neq 0, \pm \pi$ (case II) around $E(k)=0$ and $E(k)=\pm \pi$ for a driven 
	system considering slab geometry. Panels (a), and (c) indicate that the system exhibits gapless edge states (both $0$ and $\pi$) in the absence of the $d$-wave superconducting gap~($\Delta=0$). 
	These edge modes are gapped out in the presence of the superconducting term ~($\Delta \neq 0$) and are shown in panel (b) and (d). One obtains the $0$- and $\pi$-MCMs in the finite geometry 
	under OBC for panel (b) while the counter-propagating edge-modes in FFOTI intersect at $k=0,\pm \pi$. However, no $0$- and $\pi$-MCMs appear for panel (d) while the counter-propagating edge-
	modes in FFOTI intersect at $k \neq 0,\pm \pi$.
	}
	\label{DynamicalModesSchematics}
\end{figure}

The time-periodic harmonic drive, that we introduce in \Eq{drive}, cannot provide us with the $0$- and/or $\pi$-MCMs for any arbitrary choices of the parameters in the Hamiltonian $H(\vect{k})$ [\Eq{staticHam}]. To elaborate further, we first consider $\Delta=0$. In this limit, one can generate the Floquet first-order TI~(FFOTI), hosing $0$- and $\pi$-edge modes at $E(k)=0$ and $\pm \pi$ respectively, while examining the quasi-energy spectra~\cite{UmerPRB2020}. We depict the schematic representation of these edge modes in Fig.~\ref{DynamicalModesSchematics}~(a) and (c) corresponding to two cases. In case I, the counter-propagating $0$- and $\pi$-edge modes intersect with each other at $k=0$, $\pi$~[see Fig.~\ref{DynamicalModesSchematics}~(a)], whereas in case II, these edge modes intersect at other arbitrary momenta except $k = 0, \pi$~[see Fig.~\ref{DynamicalModesSchematics}~(c)] (We here use $k$ to represent the quasi-momentum in the slab geometry). These counter-propagating edge modes directly indicate the topological nature of the bulk gap at that quasi-momentum. Therefore, the generation of the FFOTI only demands a bulk topological gap at Floquet zone centre $E(\vect{k})=0$ and/or Floquet zone boundary $E(\vect{k})=\pm \pi$ irrespective of the value of the quasi-momentum $\vect{k}$ at which the topological gap is opened. We refer to this gap as the FFOTI gap. Afterward, when we consider harmonic drive on the underlying model with non-zero value of $\Delta$, these edge modes are gapped out in both the cases I and II~[see Fig.~\ref{DynamicalModesSchematics}~(b) and (d)]. However, only in case I, one can observe that the MCMs appear, when investigated employing OBC in both the directions. By contrast, case II does not 
exhibit the MCMs, within that bulk quasi-energy gap even with $\Delta \neq 0$. The above observation signals to the fact that the FFOTI gap at $\bar{\vect{k}}=(0,0)$ and $\bar{\vect{k}}=(\pi,\pi)$ can cause the FSOTSC phase to appear eventually. On the other hand, one cannot achieve FSOTSC phase once the FFOTI gap appears away from $\bar{\vect{k}}=(0,0)$ and $\bar{\vect{k}}=(\pi,\pi)$. These special momentum modes are intimately related to the details of the underlying static model. The static Hamiltonian $H(\vect{k})$ respects the spatial symmetries ${\rm S}=\left\{\mathcal{M}_x, \mathcal{M}_y, C_4, \mathcal{M}_{xy}, \mathcal{M}_{x\bar{y}} \right\}$ and the corner modes are also protected by these spatial symmetries. However, only at $\bar{\vect{k}}(=-\bar{\vect{k}}~{\rm mod}~2 \pi)$, $H(\vect{k})$ satisfy the commutation relation: $\left[H(\bar{\vect{k}}), {\rm S}\right]=0$. Thus, only the gap-closing at $\bar{\vect{k}}=(0,0)$ and $\bar{\vect{k}}=(\pi,\pi)$ gives rise to MCMs. Furthermore, the $d$-wave nature of the superconducting gap that vanishes at $\bar{\vect{k}}=(0,0)$ and $\bar{\vect{k}}=(\pi,\pi)$ is also very crucial for generating the FSOTSC phase hosting $0$ and $\pi$ MCMs. In short, the nature of the gap at $\bar{\vect{k}}=(0,0)$ and $\bar{\vect{k}}=(\pi,\pi)$ in the quasi-energy spectra for the FFOTI essentially determines the emergence of the FSOTSC phase \ie whether or not it will be observed.   

Interestingly, in the step or kick drive protocol, one can analytically derive the Floquet operator~\cite{Huang2020,HuPRL2020,ghosh2021systematic,GhoshDynamical2022}. Thus, a closed form of the 
gap closing conditions at $E(\vect{k})=0, \pi$ for $\vect{k}=(0,0)/(\pi,\pi)$ can be obtained in terms of the driving parameters. However, for a periodic sinusoidal/laser irradiation, the Floquet operator $U(\vect{k};T,0)$~[\Eq{time-evolution}] cannot be cast in a closed form. Hence, the generation of $0$- and $\pi$-MCMs are relatively cumbersome in presence of such driving protocols. 
Especially, in case of laser irradiation drive [$\vect{A}(t)=A \left( \cos \Omega t, \sin \Omega t \right)$], the quasi-momenta are replaced by $\vect{k} \rightarrow \vect{k}-\vect{A}(t)$. Thus, finding a suitable parameter space for engineering the FSOTSC can be a formidable task and will be presented elsewhere.

\subsection{Comment on comparison between the FPT and BW perturbation theory}
In the BW perturbation theory~\cite{MikamiBW2016}, one can obtain the effective Floquet Hamiltonian for a periodically driven system assuming the high-frequency limit ($\Omega \gg \gamma$). 
The BW effective Hamiltonian can be cast in the form~\cite{MikamiBW2016,Ghosh2020}
\begin{eqnarray}\label{BWeffHam}
	H_{\rm eff}^{\rm BW}(\vect{k})=H_0 + \sum_{m \neq 0} \frac{H_{-m}H_m}{m \Omega} + \mathcal{O}\left(\frac{1}{\Omega^2}\right) \ , 
\end{eqnarray}
where, $H_m$'s are defined in \Eq{FourierComp}. In Sec.~\ref{Sec:V}, we study the driven system using the FPT that assumes the strength of the drive is strong. Although, there is no limit on the frequency of the drive in the said theory. On the other hand, the BW perturbation theory contains a series of terms in the order of $1/\Omega$, and the higher-order terms in the series can be neglected in the high-frequency regime. However, with the increase in the driving amplitude, the contributions from the higher-order terms may also become more important, and curtailing the series only after the first-order 
term ($1/\Omega$-term) is not justified. Thus, to obtain a coherent picture from both the FPT and the BW perturbation theory, one needs to consider the higher-order terms in \Eq{BWeffHam}. 
Nevertheless, the BW perturbation theory cannot capture the anomalous $\pi$-modes~\cite{MikamiBW2016}. Hence, we do not provide the repeating results already investigated in the previous studies~\cite{Ghosh2020}.

\section{Summary and Conclusions}\label{Sec:VII}
To summarize, in this article, we consider a time-periodic harmonic drive protocol to generate the 2D FSOTSC anchoring both $0$- and $\pi$-MCMs. We demonstrate our results for two cases. In the first case, we start from a static model which is topologically trivial, and the driven system hosts only the $\pi$-MCMs~[see Fig.~\ref{Harmonic}~(a)]. In the second case, the starting model is topologically non-trivial and the driven system hosts both the regular $0$- and anomalous $\pi$-MCMs~[see Fig.~\ref{Harmonic}~(b)]. From the LDOS behavior, one can infer that the Majorana modes are localized at the corner of the system~[see Fig.~\ref{Harmonic}~(c)]. Further, we study the time-dynamics of the MCMs in terms of their TDOS and show that the MCMs emerge at $t<T$ and remain at $t=T$ (see Fig.~\ref{TimeDynamics}). Also, the time dynamics of the LDOS indicates sharp localization near the corners of the system for $0$- and $\pi$-MCMs when $t\sim T$.
We generalize the time-dependent dynamical nested Wilson loop technique for our case following Refs.~\cite{HuPRL2020,GhoshDynamical2022} where this has been developed for the step-drive protocol. The topological characterization of the bulk $0$ and $\pi$ gaps is carried out with the help of average quadrupolar motion (see Fig.~\ref{Invariants}). Moreover, we consider the strong driving amplitude limit and use the FPT to provide some analytical insights to our problem. We approximate the first-order term in the perturbation series and compare the FPT results with the exact numerical ones (see Fig.~\ref{PeturbationEigenvalue}). The time-evolution of the MCMs via the TDOS and the time-dependent quasi-energy gap structure (within $t\leq T$) comparison indicate that the FPT yields better results in the higher-frequency regime with that of the exact ones (see Fig.~\ref{PeturbationBulkGapTimeDynamics}), as far as $0$-MCMs are concerned. However, the FPT is unable to capture TDOS associated with the dynamical $\pi$-MCMs which we leave for future studies. The disorder analogue of the present problem can also lead to interesting outcomes about the stability of these modes. 
Also, the driving protocol and the formalism that we use in this article, can be generalized to realize the three-dimensional (3D) FSOTSC hosting 1D gapless $0\mhyphen / \pi$ Majorana hinge modes 
and 3D Floquet third-order TSC anchoring localized $0\mhyphen / \pi$ MCMs. 
\subsection*{Acknowledgments}
A.K.G. and A.S. acknowledge SAMKHYA: High-Performance Computing Facility provided by Institute of Physics, Bhubaneswar, for numerical computations.

\bibliography{bibfile}{}

\end{document}